\documentclass{PoS}

\usepackage{epsf}
\usepackage{psfig}
\usepackage{epsfig}
\usepackage{lscape}

\title{Properties and short-time evolution of nearby galaxies}

\ShortTitle{Near Galaxies}

\author{\speaker{A. Chuprikov}\\
        Astro Space Center of P.\ N.\ Lebedev Physical Institute of
Russian Academy of Sciences,Profsoyuznaya 84 / 32, 117997,
Moscow, Russia\\
        E-mail: \email{achupr@asc.rssi.ru}}

\author{I. Guirin\\
        Astro Space Center of P.\ N.\ Lebedev Physical Institute of
Russian Academy of Sciences,Profsoyuznaya 84 / 32, 117997,
Moscow, Russia\\
        E-mail: \email{igirin@asc.rssi.ru}}

\abstract{We analyze the results of processing of data of observations which had been carried out with 
the VLBA during 10 last years. All the data have been retrieved from archive of the National Radio 
Astronomy Observatory (USA NRAO, Socorro, New Mexico). Particularly, we examine data of VLBA 
observational sessions titled {\it BK068}, {\it BL111}, {\it BL137}, {\it BL149}, and {\it BD086}. 
Objects of our interest are near galaxies with z < 0.02. The radio maps of compact structure around 
the active galactic nuclei (AGN) reconstructed for two such galaxies {\it (NGC315 and 3C274)} in 
U frequency band (15 GHz) are presented. Some parameters of these sources are shown in {\it Table 1}.  
We have to perform the careful amplitude and phase calibration for all the data. Particularly, a  
correction of the delay caused by the Earth atmosphere has been made because it is necessary in this 
frequency range. Bright quasars close to the target sources (for instance, J0136+4751 and 3C279 
correspondingly) are used as atmosphere calibrators. Secondly, the Multi Frequency Synthesis (MFS) 
method (Likhachev, 2004) is used for final maps reconstruction as well as for estimation of spectral 
index values of core and jet of AGN of both objects. As a result, we can make some conclusions about 
the milliarcsecond structure of central regions of sources thanks to the using of these two methods 
of the VLBA data processing. Moreover, we can reveal some changes of this structure in 1999 - 2009, 
and estimate roughly the velocities and acceleration/deceleration values for brightest components of 
the nuclei of NGC315 and 3C274. Any polarization phenomena are not taken into account. We present 
results of processing of data of LL polarization for all the observational sessions}

\FullConference{Panoramic Radio Astronomy: Wide-field 1-2 GHz research on galaxy evolution - PRA2009\\
		 June 02 - 05 2009\\
		 Groningen, the Netherlands}

\begin{document}

\section{Results of data processing} 

The software project Astro Space Locator (ASL for Windows) (Chuprikov, 2002) has been used to process all
the data. The data processing consists of the following stages : 

$\bullet$ Amplitude calibration of all the data using GC and TY tables

$\bullet$ Single band fringe fitting (the primary phase calibration) of all the data. 
Estimation of the optimum value of solution interval

$\bullet$ Averaging of all the data over each frequency band

$\bullet$ Multi band fringe fitting of the atmosphere calibrator data. Estimation of gain values for 
every frequency and every time interval

$\bullet$ Application of gains, compensation of atmosphere delay for all the data 

$\bullet$ Self-calibration

$\bullet$ Imaging

Figure 1 demonstrates the compact structure of nuclei of 3C274 and NGC315 reconstructed according to 
the scheme above. Evolution of this structure is obvious for both sources. Table 2 and Table 3 show 
the changes of coordinates of components. These changes have the following properties : 
\begin{enumerate}
\item For 3C274\\
$\bullet$ Image consists of 3 components (A, B, and C)\\
$\bullet$ There is not any visible proper motion of the brightest component (component A)\\
$\bullet$ Velocity value of the B component is minimal between February 6, 2003 and February 5, 2007. 
This value is equal approximately to $1.5 \cdot 10^{8}$~cm/s ,
and it increases after February 5, 2007 up to $1.1 \cdot 10^{9}$~cm/s\\
$\bullet$ Velocity value of the C component also has a minimum between February 6, 2003 and February 5, 2007.
This minimum value is equal to approximately $6.1 \cdot 10^{8}$~cm/s\\ 
Again, this value increases after February 5, 2007 up to $3.6 \cdot 10^{9}$~cm/s\\

\item For NGC315\\
$\bullet$ Image consists of 2 components (A, and B)\\
$\bullet$ There is not any visible proper motion of the brightest component (component A)\\
$\bullet$ Velocity of the B component decreases between July 19, 1999 and September 12, 2008 from 
$6.5 \cdot 10^{9}$~cm/s down to $1.7 \cdot 10^{8}$~cm/s\\
\end{enumerate}

\begin{table}[red]
\caption{Redshift (Fomalont, 2000) and distance for 3C274 and NGC315}
\label{red_table}
\begin{tabular}{lcccc}
\hline
     IAU & Common & z & Distance & 1 mas corresponds to \\
     Name & Name &  & [Mpc] & linear size [cm] \\
\hline
 J1230+1223 & 3C274  & 0.004 & 14.7 & $2.20 \cdot 10^{17}$ \\
 J0057+3021 & NGC315 & 0.017 & 62.5 & $9.35 \cdot 10^{17}$ \\
\hline
\end{tabular}
\end{table}

\section{Conclusions}

We assumed that using the procedure of atmosphere delay calibration can essentially 
improve the image of source in the 2 centimetre wavelength range. The monitoring of near galaxies 
radio structure evolution for the period of approximately 10 years is possible thanks to such application 
of such procedure to VLBA data. Analysis of evolution of structure of two galaxies (3C274, and NGC315) 
demonstrates absence of superluminal motion of any component in both sources. The maximum value of component 
velocity is equal approximately to $6.5 \cdot 10^{9}$~cm/s (or $ 0.22 \cdot c$).  
Thus, we could to propose that strong synchrotron self-absorption is significant in both sources. 
Moreover, spectral index maps show a highly inverted core (spectral index is positive in core of both 
sources). This could be another indicator of self-absorption.\\
All the results presented in this paper are preliminary. It is necessary to perform similar monitoring 
of near galaxies in other frequency ranges to make final conclusions.
Procedures and techologies used during the VLBA data processing also could be very useful for 
processing of data of future Space VLBI mission titled RADIOASTRON. 

\begin{figure}[h]
\vskip2mm
\centerline{{\psfig{figure=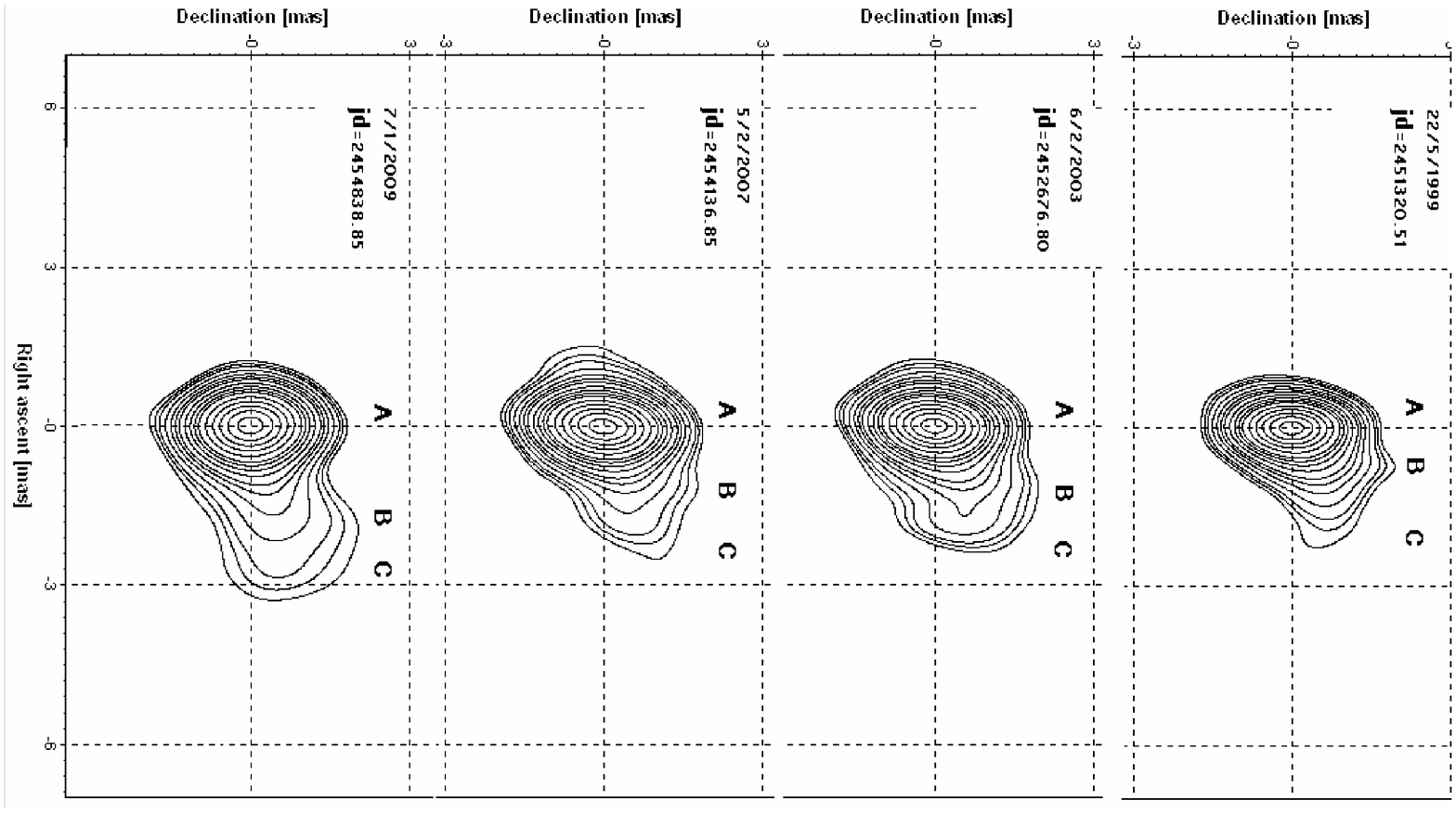,width=120truemm,angle=90,clip=}}
\hskip2mm
{\psfig{figure=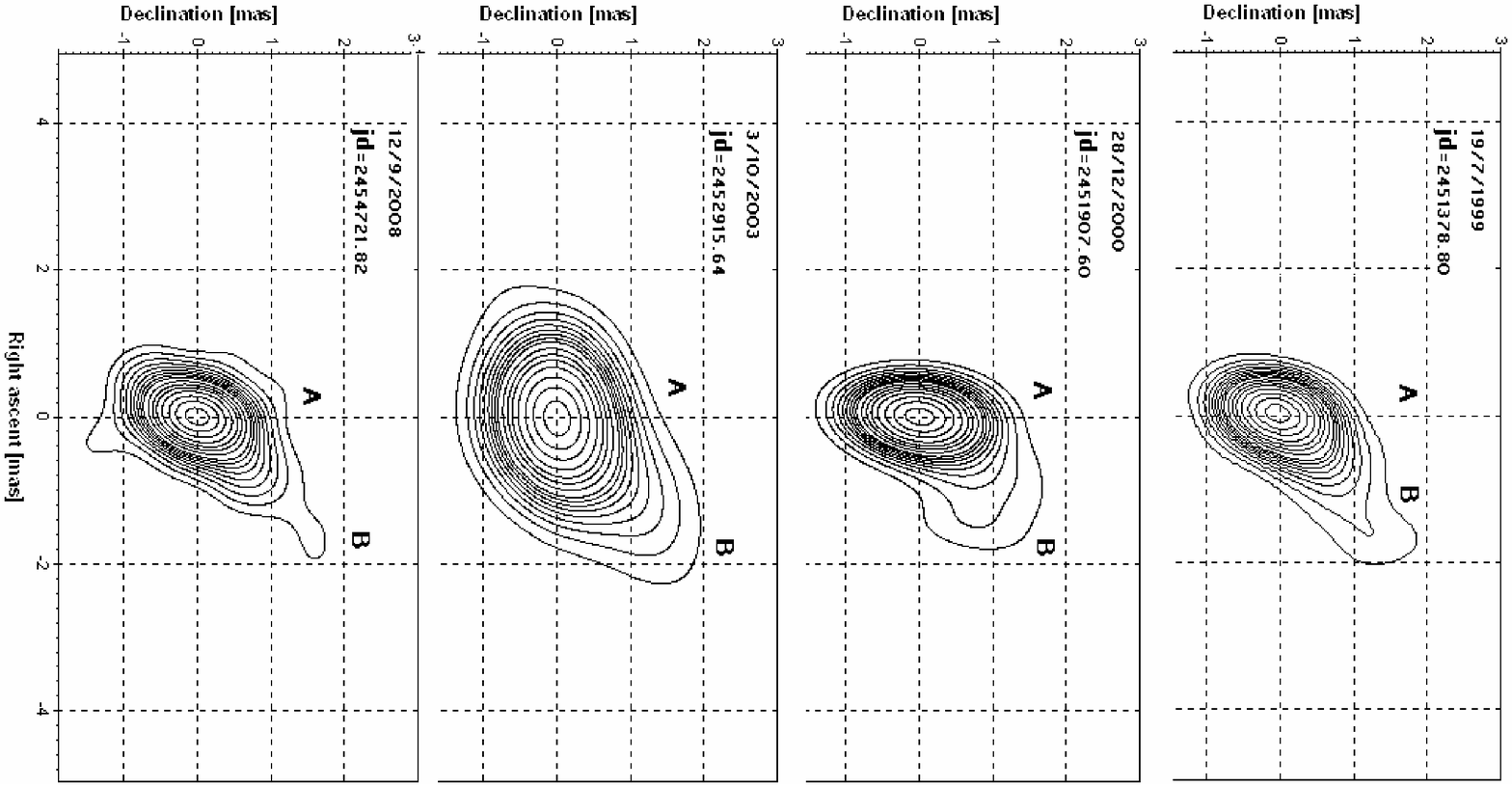,width=120truemm,angle=90,clip=}}}
\caption{Evolution of the radio structure in 1999 - 2009 for 3C274 (left), and NGC0315 (right)}
\label{fig:sim}
\vskip2mm
\end{figure}

\begin{table}[vir_a]
\caption{Relative coordinates of A, B, and C components of the central region of 3C274 in 1999 - 2009}
\label{cmp_table}
\begin{tabular}{lccccccccc}
\hline
     Date & JD & VLBA & RA of & DEC of & RA of & DEC of & RA of & DEC of \\
          & 2450000 + & Session & A (mas) & A (mas) & B (mas) & B (mas) & C (mas) & C (mas) \\
\hline
 22/05/1999 & 1320.51 & BK068 & 0 & 0 & -0.870 & 0.61 & -2.32 & 0.64 \\
 06/02/2003 & 2676.80 & BL111 & 0 & 0 & -1.552 & 0.59 & -2.52 & 1.01 \\
 05/02/2007 & 4136.85 & BL137 & 0 & 0 & -1.637 & 0.57 & -2.40 & 0.68 \\
 07/01/2009 & 4838.85 & BL149 & 0 & 0 & -1.900 & 0.40 & -3.35 & 0.98 \\
\hline
\end{tabular}
\end{table}

\begin{table}[ngc315]
\caption{Relative coordinates of A and B components of the central region of NGC315 in 1999 - 2008}
\label{cmp1_table}
\begin{tabular}{lccccccc}
\hline
     Date & JD & VLBA & RA of & DEC of & RA of & DEC of \\
          & 2450000 + & Session & A (mas) & A (mas) & B (mas) & B (mas) \\
\hline
 19/07/1999 & 1378.97 & BK068 & 0 & 0 & -1.618 & 1.196  \\
 28/12/2000 & 1907.38 & BK068 & 0 & 0 & -1.437 & 0.930  \\
 03/10/2003 & 2915.65 & BD086 & 0 & 0 & -1.525 & 1.486  \\
 12/09/2008 & 4721.82 & BL149 & 0 & 0 & -1.500 & 1.500  \\
\hline
\end{tabular}
\end{table}

\end{document}